\documentclass[%
 reprint,
 sd,%
 amsmath,amssymb,
 aip,
apl,
]{revtex4-2}

\usepackage{graphicx}
\usepackage{dcolumn}
\usepackage{bm}

\usepackage{subfigure}

\begin{document}

\preprint{APS/123-QED}

\title{Rapidly Switchable X-ray Angular Momentum from a Free Electron Laser Oscillator}

\author{Nanshun Huang}
\affiliation{%
Zhangjiang Laboratory, Shanghai 201210, China
}%

 
\author{Haixiao Deng}%
 \email{denghx@sari.ac.cn}
\affiliation{%
Shanghai Advanced Research Institute, Chinese Academy of Sciences, Shanghai 201210, China
}%

\date{\today}

\begin{abstract}

X-ray vortices carrying tunable Orbital Angular Momentum (OAM) are an emerging tool for X-ray characterization technology.
However, in contrast to the generation of vortex beams in the visible wavelength region, the generation of X-ray vortices in a controlled manner has proved challenging. Here, we overcome this challenge using an X-ray free-electron laser oscillator (XFELO), which can produce intense coherent X-rays with  switchable OAM. Using the pinhole mirror in an XFELO, this scheme adjusts only the kinetic energy of the electron beam to produce vortex beams that can be programmed to dynamically change between different OAM modes without the need for additional optical elements. With the nominal parameters of currently constructing 1~MHz repetition rate facility (i.e. SHINE), the active formation of the OAM modes of $l = \pm 1$ and $l = \pm 2$ and the rapid switching between them by detuning the electron beam energy of the XFELO are numerically illustrated. The real-time switching can be achieved within 200 $\mu$s, while the output pulse energy can reach the 100 $\mu$J level. 


\end{abstract}

\maketitle

Spin angular momentum (SAM) and orbital angular momentum (OAM) are two distinct properties of light. The SAM of light take values only of $+1$ or $-1$, referring to the left and right circular polarization states respectively. Besides the SAM, photon beams with spiral wavefronts, characterized by a transverse phase structure of $\exp(\pm il\phi)$, can carry the OAM equivalent to $l\hbar$ per photon (where $\phi$ is the azimuthal angle, and the integer $l$ is known as the vortex order or topological charge)~\cite{Allen1992,rubinsztein-dunlop_roadmap_2016}. Since the OAM can be much greater than the SAM by tuning the topological charge, OAM light beams have been exploited in various applications~\cite{shen_optical_2019}, including micromanipulation, imaging, high-capacity communications, and quantum entanglement. A recent study demonstrated time-dependent OAM beams in extreme ultraviolet light that result in a self-torque~\cite{rego_generation_2019}. 

\begin{figure*}[!htb]
  \centering
  \subfigure{\includegraphics*[width=420pt]{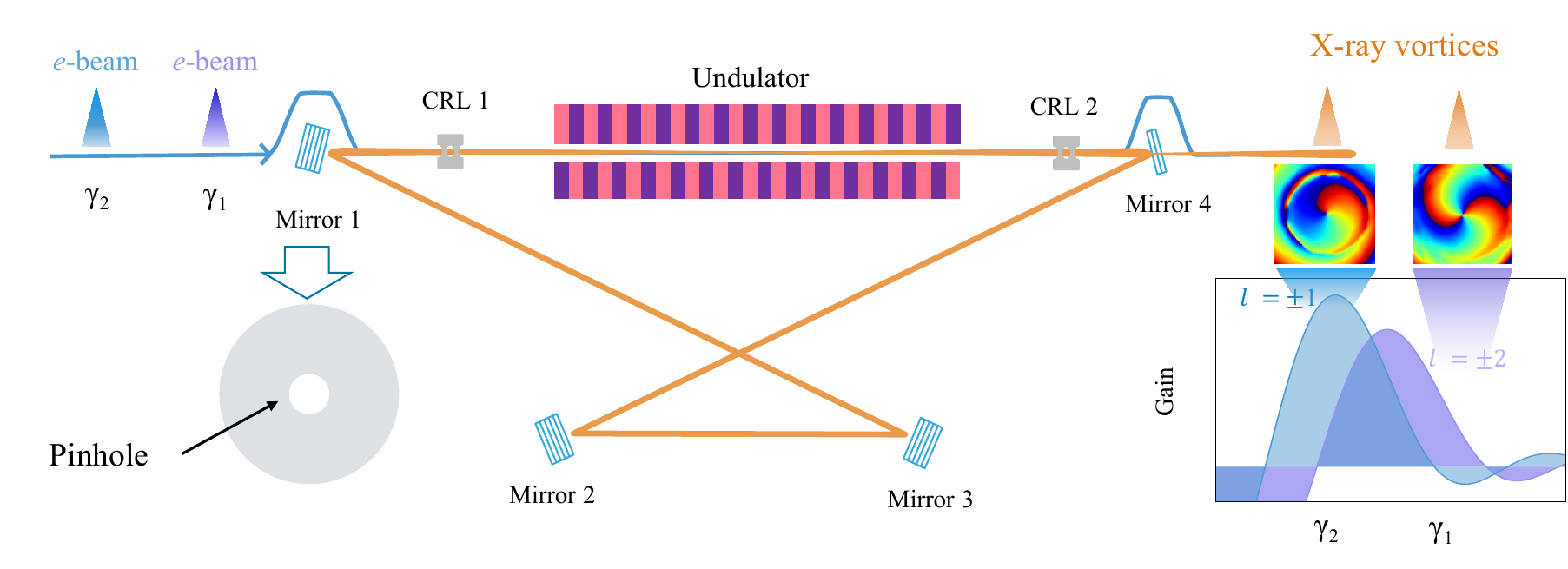}}
  \caption{General setup for XFELO optical vortex generation and OAM mode switching. Controlling the electron beam energy together with the ultra-narrow bandwidth, enables the manipulation of the OAM mode orders}
  \label{fig:scheme}
\end{figure*}

However, the exploitation of X-ray OAM beams is a relatively novel domain. OAM X-ray beams hold the potential to probe material properties in a way that is inaccessible using traditional X-ray, such as probing a novel type of phase dichroism in resonant diffraction from an artificial antiferromagne~\cite{mccarter_antiferromagnetic_2022}, inducing charge current loops in fullerenes controlled by tuning the topological charge~\cite{watzel_optical_2016}, and enhancing microscopy~\cite{kohmura_x-ray_2020} and X-ray imaging~\cite{pancaldi_high-resolution_2023}. Very recently, helicity-dependent spectroscopic signal was experimentally measured from a hard X-ray OAM beam at the iron (Fe) K edge (7.1 keV, 1.7 \AA)~\cite{rouxel_hard_2022}. Overall, interest in optical vortices in the X-ray regime is increasing.


Most OAM-based proposals depend on the difference in the response of matter to the topological charge $l$ of the vortex~\cite{pancaldi_high-resolution_2023,rouxel_hard_2022, ye_probing_2019}. Rapid switching between OAM modes is highly desirable to fully exploit the benefits of those OAM-based methods. This requires a high-brightness X-ray source capable of producing tunable OAM modes. In contrast to the generation of vortex beams at ultraviolet wavelengths~\cite{gauthier_tunable_2017,heras_extreme-ultraviolet_2022}, the generation of intense adjustable X-ray vortices remains challenging and has been extensively investigated. On timescales of tens of minutes, it was recently demonstrated that the OAM of the X-ray beam from an artificial spin ice can be switched by increasing the temperature from 270 K to 380 K or by applying a magnetic field~\cite{woods_switchable_2021}. However, adjusting the X-ray OAM modes via a programed manner at high switching rates is a challenge that has yet to be addressed. Considerable attention has been given to their efficient creation and manipulation.

X-ray free-electron lasers (XFELs) are new generation accelerator-based light sources with unprecedented brightness at the angstrom wavelength and femtosecond time scales~\cite{pellegrini_physics_2016, huang_features_2021,FLASH_2007,LCLS_2009,SACLA_2012,PALXFEL2017,swissFEL2020,EXFEL2020,FERMIFEL2012,SXFEL_2022_EEHC}. With their unique features, high-brightness FEL X-ray optical vortices can lead to new research fields. Therefore, several FEL schemes have been studied to generate vortex beams. In addition to direct conversion approaches~\cite{rebernik_ribic_extreme-ultraviolet_2017,loetgering_generation_2020}, these schemes exploits a helical undulator with harmonic emission or harmonic modulation~\cite{hemsing_helical_2009,hemsing_generating_2011,hemsing_echo-enabled_2012}. However, to adjust the OAM in these schemes, the operating harmonics of the helical undulator or transverse mode of the seed laser must be changed mechanically. To date, no methods has been proposed for implementing rapid and easy OAM switching using these schemes.

In this work, we propose a method for generating intense X-ray beams carrying an OAM capable of rapid switching. It is based on an X-ray FEL oscillator (XFELO) with a pinhole mirror, which consists of an optical cavity (to circulate X-ray pulses) and an undulator (to provide gain). The XFELO is a promising candidate for generating intense, fully coherent X-ray pulses at high repetition rates~\cite{XFELO_kim_2008,XFELO_harmonic_2012_deng,Marcus_RAFEL_2020,EXFEL_XFELO_cavity_2023_prab}. The scheme layout is shown in Fig.~\ref{fig:scheme}. In the proposed method, an XFELO can directly generate X-ray vortices whose topological charge can be adjusted by changing only the electron beam energy, thereby allowing for very fast OAM switching and programming without changing the mechanical setup. The underlying principle of modifying the OAM order through electron beam energy adjustment is rooted in the FEL nature, where the gain spectrum for various transverse modes experiences a shift due to mode-dependent Gouy phase detuning, which was initially introduced for the direct generation of OAM beams~\cite{huang_generating_2021}.



The impact of the Gouy phase on FEL amplification was previously studied ~\cite{saldin_linear_1993,hemsing_longitudinal_2008,pellegrini_physics_2016,huang_generating_2021}. The presence of the Gouy phase causes additional phase slippage when the radiation field passes through the focus. This additional phase slippage increases the phase speed of light, and a faster electron velocity is required to maintain resonance. Therefore, the optimal detuning point is shifted to higher beam energies~\cite{pellegrini_physics_2016}. Thus, for certain FEL parameters, it is possible to detune such that the OAM modes have maximum gain. 

This effect can be derived from the change in the FEL resonance condition. The paraxial beam $E(r,\phi,z)$ can be expanded in a series of Laguerre--Gauss (LG) modes as follows:
\begin{equation}
  \mathrm{LG}_{p}^{l}(r, \phi, z)=A_{p}^{|l|}(r, \phi, z) e^{i l \phi} e^{i \varphi(z)} 
\end{equation}
with 
$$
\varphi(z) = (1+2p+|l|) \arctan(\frac{z}{z_{\mathrm{R}}}), 
$$
where $\varphi$ is the Gouy phase, $A_{p}^{l}$ is the complex amplitude, $l$ denotes the azimuthal mode number, and $p$ is the radial mode number. $z_{\mathrm{R}} = {\pi w^2_0}/{\lambda}$ denotes the Rayleigh range with the wavelength $\lambda$ and beam waist $w_0$.


Then, in the basic FEL equations, the relative electron phase associated with the radiation and the undulator fields is given by $\theta = (k+k_u)z-\omega t - \varphi(z) + l\phi$. Effective energy exchange between the electron and radiation ocurrs if $d \theta/dt = 0$. Then, the resonance energy can be obtained as~\cite{kim_synchrotron_2017}
\begin{equation}
  \frac{\gamma_R^2}{\gamma_{R0}^2} =  \frac{1}{1-\frac{1}{k_u}\frac{d\varphi}{dz}},
\end{equation}
where $\gamma_{R0} = \sqrt{\lambda_u (1+K^2/2)/2\lambda} $ denotes the original FEL resonance energy with the radiation wavelength $\lambda$, undulator parameter $K$, and undulator period $\lambda_u$. $k_u = 2 \pi /\lambda_u$ denotes undulator wavenumber. Owing to the presence of $z$-dependent phase $\varphi(z)$, the resonance energy for the individual $l$ modes is different. And the change in resonance energy $\Delta \gamma_{R} = \gamma_R - \gamma_{R0}$ increases with $N=1+2p+|l|$. As a result, energy detuning can be controlled to maximize the gain of a specific mode. Through this approach, the OAM modes can be generated and amplified in an XFELO. 


\begin{figure*}[!htb]
  \centering
  \subfigure{\includegraphics*[width=460pt]{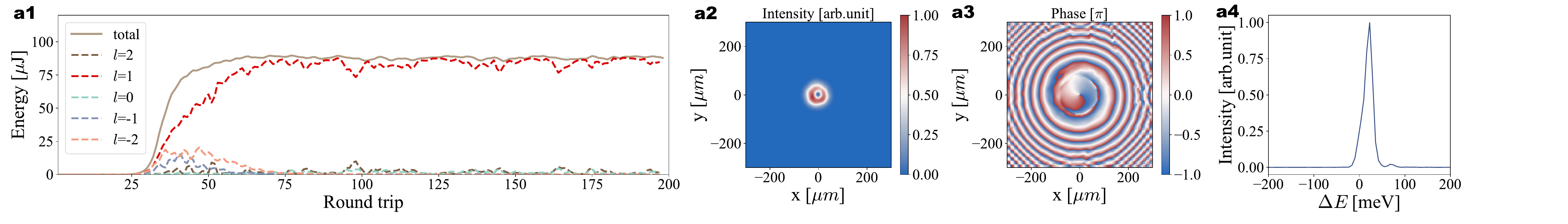}}
  \subfigure{\includegraphics*[width=460pt]{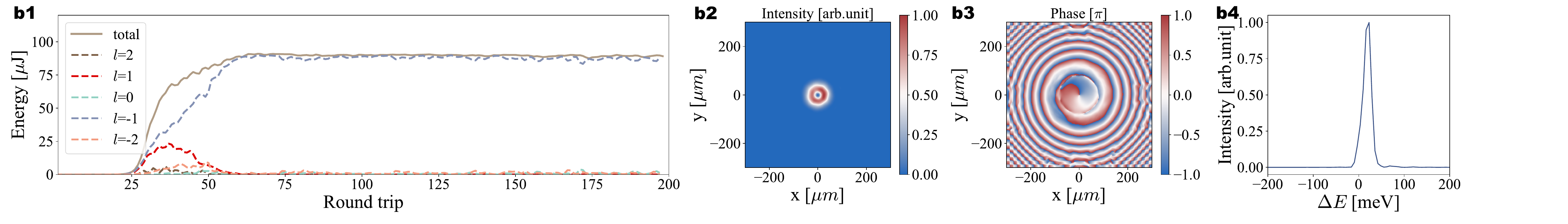}}
  \subfigure{\includegraphics*[width=460pt]{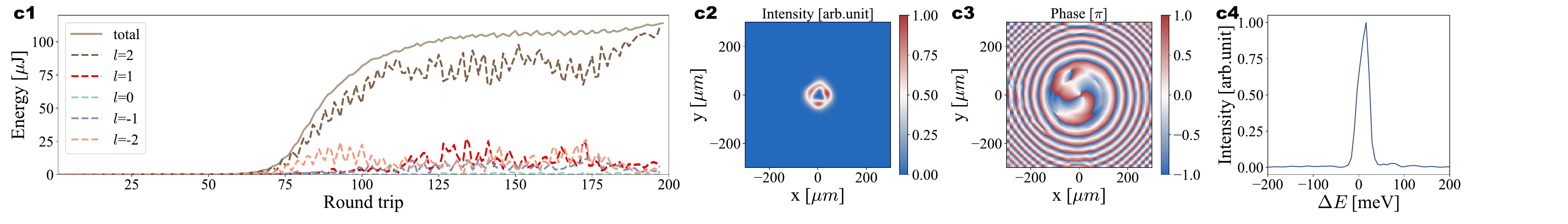}}
  \subfigure{\includegraphics*[width=460pt]{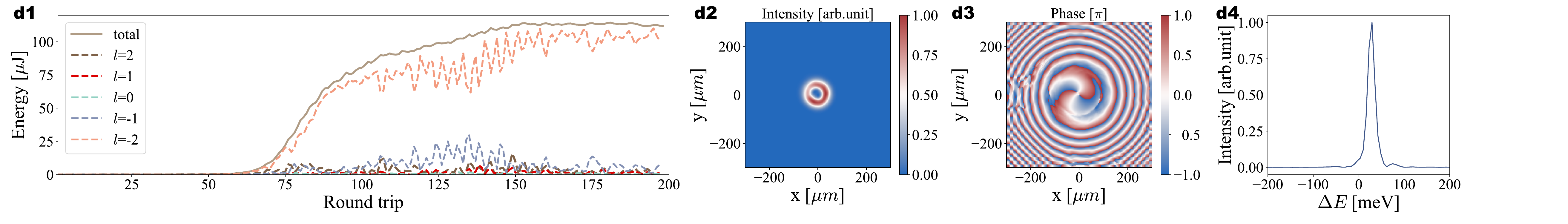}}
  \caption{Characterization of the XFELO in the OAM modes. (a), (b), (c), and (d) show the X-ray OAM beam with $l=1$, $l=-1$, $l=2$, and $l=-2$, respectively. The numbers indicate the gain curve, amplitude, phase and spectrum of the OAM beam respectively.}
  \label{fig:OAMs_s2e}
\end{figure*}


It should be noted that the $l$ and $-l$ modes are not degenerate in XFELO. This occurs because the interaction between the light and electrons leads to a micro-bunching structure in a FEL, which is imprinted by the wavefront. Owing to orthogonality ($\int \exp[i(l-k)\phi] \propto \delta_{l,k}$), the micro-bunching structures formed by different $l$ modes are not degenerate. This leads to the robust generation of a single OAM mode. However, it's worth mentioning that degenerate states might also appear when the micro-bunching is not strong enough, specifically during linear gain regions. Moreover, a detailed analysis can be found in Supplementaries.

It should also be remarked that the XFELO cannot distinguish between a change in $l$ and a change in $p$ if $N$ doesn't change. For example, if $N=1+2p+|l|=3$, the three modes exhibit optimal detuning: $p=1, l=0$ and $p=0, l=\pm 2$. In particular, undesirable $p>0$ modes are amplified. To spatially suppress the $p>0$ modes, a pinhole mirror can be used, which introduces additional loss into these modes. Therefore, by employing a pinhole, when changing the detuning, only $l$ changes, without exciting higher-order modes with $p>0$.


As analyzed above, lasing and switching in the OAM modes are realized by maximizing the gain via careful tuning of the electron beam energy and the use of a pinhole mirror. Such a small energy change is easily realized by programming the low-level RF system of the last accelerating module, as designed in the European XFEL~\cite{EXFEL2020}. In addition, the pinhole diamond mirror has already been proposed as an X-ray regenerative amplifier free-electron laser~\cite{Marcus_RAFEL_2020}.

We explored OAM mode switching with three-dimensional numerical simulations within the context of the Shanghai High-Repetition-Rate XFEL and Extreme Light Facility (SHINE)~\cite{huang_ming_2023}, which is the first hard X-ray FEL facility in China (currently under construction). The primary parameters are listed in Table~\ref{tab:SHINE_para}. To demonstrate the robustness of the OAM XFELO, a shot-to-shot jitter was applied, where the angular fluctuation of the mirrors was set at 15~nrad (RMS).


\begin{table}[!htb]
  \centering
\caption{\label{tab:SHINE_para}%
Main Parameters
}
\begin{tabular}{lcr}
\hline
Beam Energy                   & 8~GeV    \\ 
Relative Energy Spread        & 0.01\%     \\ 
Normalized Emittance          & 0.45~mm$\cdot$mrad    \\
Peak Current                  & 750~A      \\
Undulator Period Length       & 26~mm    \\
Undulator Segment Length      & 4~m \\
Photon Energy                 & 6.96~keV   \\
Mirror Material               & Diamond    \\
Bragg Mirror Reflectivity     & 85\% \\

Hole Diameter         & $ 70 ~\mu$m      \\

Bunch Transverse Size         & $\sim 37 ~\mu$m \\
CRL Focal Length               & $ 54.7$ m     \\

\hline
\end{tabular}
\end{table}



\begin{figure*}
  \centering
  \subfigure{\includegraphics*[width=460pt]{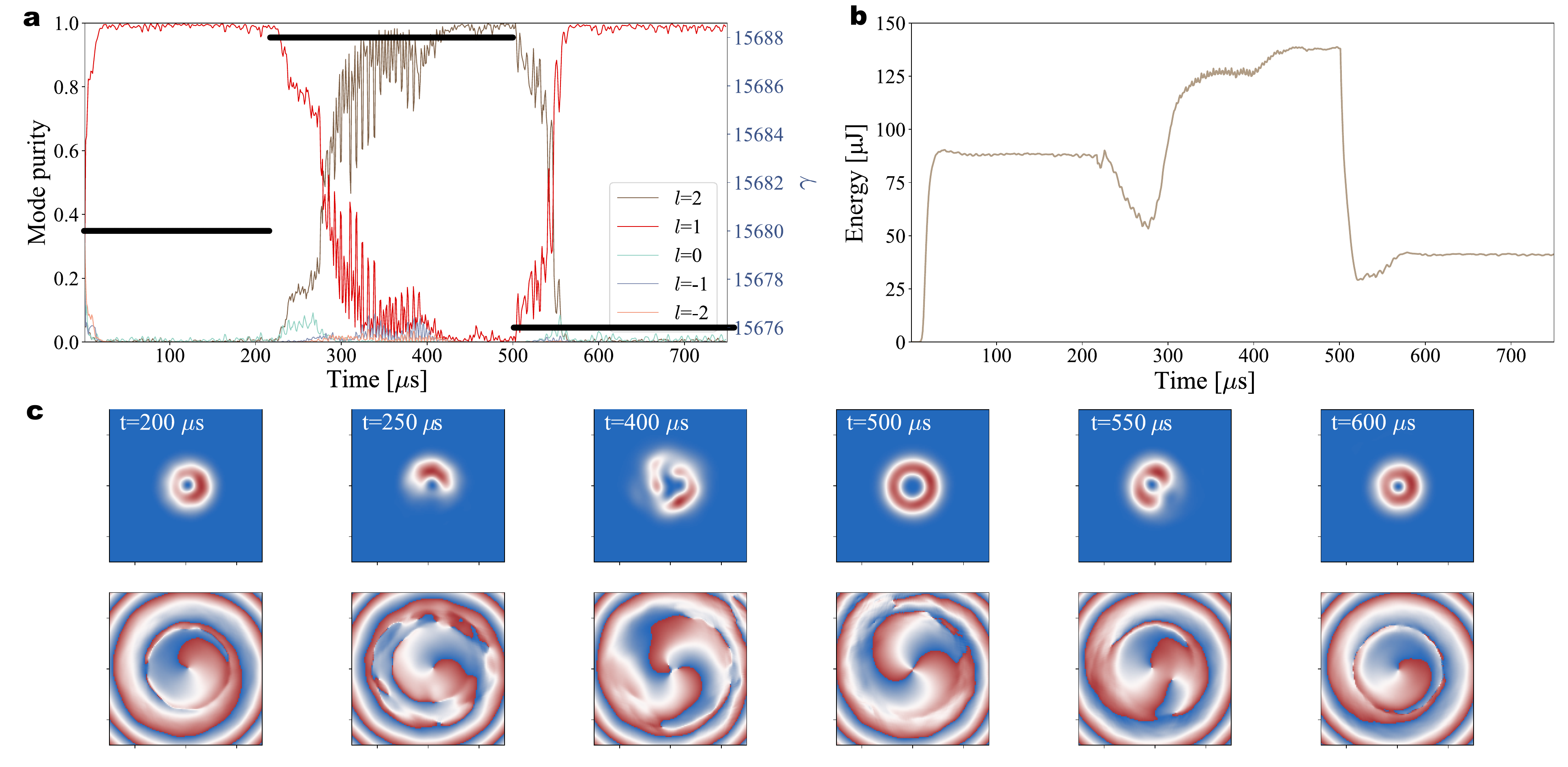}}
  \caption{Top panel: (a) mode decomposition and programed electron beam energies, (b) power evolution. Right panel: (c) snapshot images of the transverse intensity and phase distributions, showing OAM state switching between $|l|=1$ and $|l|=2$. In this case, the switching time is 200 $\mu$s.}
  \label{fig:OAMs_switching}
\end{figure*}



\paragraph{Direct lasing of different OAM modes}

To investigate the performance of the XFELO in different OAM modes, electron beams with different energies were applied (while the XFELO configuration was fixed). Two beam energies were used. The first beam energy was $\gamma=15680$, corresponding to $\frac{\delta \gamma}{\gamma_{R0}}=0.055\%$, and was used to generate the $|l|=1$ mode. The second one was $\gamma=15687$, corresponding to $\frac{\delta \gamma}{\gamma_{R0}}=0.1\%$, and was used to generate the $|l|=2$ mode. The results are presented in Fig.~\ref{fig:OAMs_s2e}.

Fig.~\ref{fig:OAMs_s2e}~(a) and (b) depicts the cavity output and mode decomposition for the $|l|=1$ charged vortex case. The pulse energy output from the optical cavity coupling reaches 100~$\mu$J. Typical transverse intensity and phase distributions at the exit of the undulator are shown in Fig.~\ref{fig:OAMs_s2e}~(a2, b2) and Fig.~\ref{fig:OAMs_s2e}~(a3, b3), respectively. The hollow intensity distribution, which is a characteristic of an optical vortex, can be found. The phase distribution represents the $l=\pm1, p=0$ LG mode, confirming the unit charge of the vortex beam.


Fig.~\ref{fig:OAMs_s2e}~(c) and (d) depicts the XFELO performance in the OAM mode of $|l|=2$. The pulse energy coupling out from the resonant cavity is approximately 120 $\mu$J at saturation. Because of the relatively low gain, the radiation in the $|l|=2$ modes requires more round trips to reach saturation. The lower gain is due to the weaker transverse coupling and the larger emission angle of larger $|l|$ modes. In addition, the lower single-pass gain in the $|l|=2$ modes leads to more severe mode competition. Thus, it takes longer to establish the dominant mode, compared to the $|l|=1$ case. Fig.~\ref{fig:OAMs_s2e}~(c2, d2) and Fig.~\ref{fig:OAMs_s2e}~(c3, d3) show the The transverse intensity and phase distributions, respectively, clearly proving that the vortex beam carries a topological charge of $|l|=2$. In both cases, high-purity optical vortices are obtained. In addition, distinctive narrow bandwidths are obtained from the these cases, as shown in right column of  Fig.~\ref{fig:OAMs_s2e}.




\paragraph{Dynamic switching of OAM modes}

In addition to the direct lasing of the different vortex beams for macro-pulse facilities, real-time switching between the OAM modes can be achieved for continuous-wave facilities. Fig.~\ref{fig:OAMs_switching} shows the cases for electron $\gamma$ values from 15676 to 15688, corresponding to OAM mode switching between $|l|=1$ and $|l|=2$.


The left panel in Fig.~\ref{fig:OAMs_switching} shows the (a) mode purity and associated electron beam energy, (b) power evolution, and (c) snapshots of intensity and phase distribution. The initial beam energy is $\gamma = 15680$, and the XFELO operates in a vortex state with the $l=1$ mode. As the electron beam energy is varied to $\gamma = 15688$, the OAM mode shifts from $l=1$ to $l=2$, shown in Fig.~\ref{fig:OAMs_switching}~(a). Subsequently, the system is switched to the emitted vortex state of $l=1$ by tuning the electron beam energy to $\gamma = 15676$. The snapshots of intensity and phase distribution in plot (c) of Fig.~\ref{fig:OAMs_switching}, collected at different times, clearly show the OAM state switching between $l=1$ and $l=2$. In this case, the switching time is determined as 200 $\mu$s. These results also suggest that X-ray OAM can be programmed by the electron beam energy.

It is interesting to observe that during the switch from the $l=1$ mode, the field is shifted, then causing two vortices with a topological charge of 1 each to merge into a single phase singularity with $l=2$. Conversely, when switching from $l=2$ to $l=1$, the single phase singularity splits into two vortices, each with a topological charge of 1. This phenomenon occurs in the interaction among modes with $|l| > 1$ due to the symmetry in phase around the axis. The dominant mode moves towards the center, causing the fusion and fragmentation of phase singularities and their relocation onto the axis. Therefore, the switching of $l$ always commences from the adjacent mode, implying that the switch could follow paths such as$ l = 2 \rightarrow  $ or $ l=-1 \rightarrow -2$, and so on. It is crucial to note that during these switch, the sign does not alter.



In conclusion, we presented a scheme for efficiently generating and manipulating X-ray vortices in an XFELO. This scheme adjusts only the electron energy, thereby allowing a fast switching between different OAM modes. To the best of our knowledge, this is the first study to propose a method for realizing fast OAM state switching in the X-ray region. Therefore, our switching rate (~kHz level in the context of SHINE) is expected to be two to three orders of magnitude faster than those of previous optical approaches. These findings represent a essential step toward the realization of programmable X-ray OAM beams. More extensively, the proposed approach holds promise for at-source solutions for a wide range of applications involving vortex beams in different areas of fundamental and applied physics, ranging from super-resolution imaging to material characterization, and from quantum optics to optical manipulation.

\acknowledgments

We thank Zihan Zhu for providing the start-to-end simulated electron bunch. This work was supported by the CAS Project for Young Scientists in Basic Research (YSBR-042), the National Natural Science Foundation of China (12125508, 11935020),Program of Shanghai Academic/Technology Research Leader (21XD1404100), and Shanghai Pilot Program for Basic Research - Chinese Academy of Sciences, Shanghai Branch (JCYJ-SHFY-2021-010). 

The authors declare no competing interests. The code used in this study is available from the corresponding authors upon request.

\end{document}